\documentclass[showpacs,aps,graphicx,twocolumn]{revtex4}%
\usepackage{graphicx}

\begin{document}
\title{Robustness of two-way quantum communication protocols against Trojan horse attack}
\author{ Fu-Guo Deng,$^{1,2,3}$  Ping Zhou,$^{1,2}$  Xi-Han Li,$^{1,2}$ Chun-Yan Li,$^{1,2}$
and Hong-Yu Zhou$^{1,2,3}$}
\address{$^1$ The Key Laboratory of Beam Technology and Material
Modification of Ministry of Education, Beijing Normal University,
Beijing 100875,
China\\
$^2$ Institute of Low Energy Nuclear Physics, and Department of
Material Science and Engineering, Beijing Normal University,
Beijing 100875, China\\
$^3$ Beijing Radiation Center, Beijing 100875,  China}
\date{\today }

\begin{abstract}
We discuss the robustness of two-way quantum communication
protocols against Trojan horse attack and introduce a novel
attack, delay-photon Trojan horse attack. Moreover, we present a
practical way for two-way quantum communication protocols to
prevent the eavesdropper from stealing the information transmitted
with Trojan horse attacks. It means that two-way quantum
communication protocols is also secure in a practical application.
\end{abstract}
\pacs{03.67.Hk, 03.67.Dd} \maketitle


Quantum communication supplies some novel ways for the
transmission of message, such as quantum key distribution (QKD)
\cite{BB84,Gisinqkd,longqkd,CORE,BidQKD,LM,QDKD}, quantum secure
direct communication (QSDC)
\cite{two-step,QOTP,QPA,Wangc,bf,cai2,cai,yan,gao,zhangzj},
quantum secret sharing (QSS)
\cite{HBB99,KKI,deng2005,COREQSS,longqss,guoqss,Peng,dengmQSTS,dengpra,zhanglm},
and so on. QKD provides a secure way for creating a private key
with which two remote parties, say Alice and Bob, can communicate
in an unconditionally secure way even though an eavesdropper, Eve
is monitoring the channel. After Bennett and Brassard published
their pioneering work \cite{BB84} in 1984, called BB84, QKD
attracts a lot of attentions
\cite{BB84,Gisinqkd,longqkd,CORE,BidQKD,LM,QDKD} and becomes one
of the most mature applications of quantum information.

QSS and QSDC are two important branches of quantum communication,
and have been developing quickly in recent years. QSS is the
generalization of classical secret sharing \cite{Blakley} into
quantum scenario and supplies a secure way for sharing both a
piece of classical information
\cite{HBB99,KKI,deng2005,COREQSS,longqss,guoqss} and quantum
information \cite{Peng,dengmQSTS,dengpra,zhanglm}. The gents can
obtain the message sent by the sender only when they cooperate,
otherwise they can get nothing. QSDC is used to transmit the
secret message directly \cite{two-step,QOTP,QPA,Wangc,bf,cai2,cai}
without creating a private key and then encrypting the message.

Recently, there are some two-way protocols proposed for QKD
\cite{BidQKD,LM,QDKD}, QSS \cite{deng2005,zhanglm}, and QSDC
\cite{two-step,QOTP,QPA,Wangc,bf,cai2,cai}. Although there are
differences among particular quantum communication protocols,
almost all of them include the following procedures
\cite{BidQKD,LM,QDKD,deng2005,zhanglm,two-step,QOTP,QPA,Wangc,bf,cai2,cai}.
First, the receiver of information, say Bob, prepares the quantum
signal randomly in one of some nonorthogonal states or a mixed
state, and sends it to the sender Alice. Alice chooses one of two
modes, checking mode and coding mode, to deal with the quantum
signal. If Alice chooses the checking mode, she will obtain a
sample for eavesdropping check with the measurement on the signal,
otherwise she operates the signal with local unitary operations
and sends it back to Bob. Bob measures the signal and gets the
information transmitted by Alice. There are at least two
transmissions of the quantum signal, i.e., from Bob to Alice, and
from Alice to Bob. These two-way quantum communication protocols
can be attacked with a Trojan horse if the two parties use only a
simple way for eavesdropping check. On the other hand, this class
of attacks can be detected with a little of modification in the
eavesdropping checks. In this paper, we will discuss three types
of Trojan horse attacks on two-way quantum communication protocols
and present a way for improving their security against those
attacks with a photon number splitter (PNS: 50/50) and a
wavelength filter.

The typical two-way quantum key distribution protocol can be
described as follows with polarized single photons
\cite{BidQKD,LM}.

(1) Bob prepares a polarized single photon in one of the four
states $\{\vert +z\rangle, \vert -z\rangle, \vert +x\rangle, \vert
-x\rangle\}$ randomly. Here
\begin{eqnarray}
\vert +z\rangle=\vert 0\rangle,\\
\vert -z\rangle=\vert 1\rangle,\\
\vert +x\rangle =\frac{1}{2}(\vert 0\rangle + \vert 1\rangle),\\
\vert -x\rangle =\frac{1}{2}(\vert 0\rangle - \vert 1\rangle),
\end{eqnarray}
and $\{\vert +z\rangle, \vert -z\rangle\}$ are the two eigenstates
of $\sigma_z$, and $\{\vert +x\rangle, \vert -x\rangle\}$ are
those of $\sigma_x$.

(2) Bob sends the photon to Alice, and Alice chooses one of the
two modes to deal with it. If she chooses checking mode, Alice
measures the photon with one of the two measuring bases (MBs),
$\sigma_z$ and $\sigma_x$, randomly. If she chooses the coding
mode, she operates the photon with one of the two unitary
operations $I$ and $U$. Here $I=\vert 0\rangle\langle 0\vert +
\vert 1\rangle\langle 1\vert$ is the identity matrix and $U=\vert
0\rangle\langle 1\vert - \vert 1\rangle\langle 0\vert$. The nice
feature of the $U$ operation \cite{QOTP,BidQKD} is that it flips
the state in both measuring bases, i,e,
\begin{eqnarray}
U\{\vert +z\rangle, \vert -z\rangle\}=\{-\vert -z\rangle, \vert
+z\rangle\}, \\
U\{\vert +x\rangle, \vert -x\rangle\}=\{\vert -x\rangle, -\vert
+x\rangle\}.
\end{eqnarray}

(3) Alice sends the photon operated to Bob, and Bob measures it
with the same MB as that he prepares it.

(4) Alice and Bob repeat the process until they obtain enough key
bits.

(5) Alice and Bob analyze the error rate of the samples obtained
with the checking mode. Moreover they pick out randomly a subset
of the outcomes obtained by Bob for eavesdropping check.

(6) If they can determine that the quantum channel is secure, they
do error correction and privacy amplification on the outcomes to
distill a private key.

In fact, this QKD protocol is equivalent to two BB84 QKD protocol
if the eavesdropper Eve does not attack it with a Trojan horse.
The latter is proven unconditionally secure \cite{BB842}. If the
QKD protocol is robust against Trojan horse attacks, it is also
secure.

Now, let us introduce the three types of Trojan horse attacks. The
first one is the general Trojan horse attack introduced in Ref.
\cite{Gisinqkd}. That is, Eve sends a light pulse to Alice, same
as Bob. The second is a new one, the delay-photon Trojan horse
attack. In detail, Eve intercepts the signal transmitted from Bob
to Alice, and then inserts a fake photon in the signal with a
delay time, shorter than the time windows \cite{Gisinqkd}. In this
way, Alice cannot detect this fake photon as it does not click
Alice's detector. After the operation done by Alice, Eve
intercepts the signal again and separates the fake photon. She can
get the full information about Alice's operation with measurement.
The third Trojan horse attack is the invisible photon attack
proposed by Cai recently \cite{caiattack}. Its main idea is that
Eve inserts an invisible photon in each signal prepared by Bob and
sends it to Alice. As Alice's detector cannot click this photon
and performs an unitary operation on each signal, Eve can steal
the information about Alice's operation by means that she
intercepts the signal operated and separates the invisible photon
from each signal. With the measurement on the invisible photon,
Eve can read out Alice's information. Its implement may be resort
to the second attack strategy as it is necessary for Eve to
separate the invisible photon from the signal without destroying
the original photon.

In essence, the security of quantum communication protocol comes
from the fact that the authorized users can detect the
eavesdropper with measurements on some samples. For the third
Trojan horse attack proposed by Cai \cite{caiattack}, Alice needs
only to add a wavelength filter \cite{Gisinqkd} on each signal
before she deals with it (i.e. coding or measuring it). In
practical quantum communication, Alice and Bob should exploit a
wavelength filter to filtering the light from background, in
particular in free space. So there is no problem for the users to
deal with this attack.

For the delay-photon Trojan horse attack, Alice should use a PNS
to divide each sample signal into two pieces and measure them with
two MBs. If there is only one photon in the original signal, Alice
can only get one outcome, otherwise she will obtain two outcomes.
In this way, Alice can improve the security of two-way quantum
communication protocols against the Trojan horse attacks.
Obviously, the method is also efficient for Alice to avoid the
first Trojan horse attack. In practical, photon number splitting
technique is not easy to be implemented with current technology
\cite{Gisinqkd}, a photon beam splitter (PBS: 50/50) which is not
difficult to be made can be used to replace the PNS . If the time
windows of the two single-photon devices is long enough, Alice can
detect the eavesdropping with a multi-photon fake signal.

In summary, a PBS and a wavelength filter can be used to avoid the
two-way quantum key distribution protocol against Trojan horse
attacks. The same result can be drawn for the other two-way
quantum communication protocols, such as QSS
\cite{deng2005,zhanglm} and QSDC
\cite{two-step,QOTP,QPA,Wangc,bf,cai2,cai} protocols.


This work is supported by the National Natural Science Foundation
of China under Grant Nos. 10447106, 10435020, 10254002, and
A0325401, and Beijing Education Committee under Grant No.
XK100270454.


\end{document}